\documentclass[aps,prb,preprint,preprintnumbers,showpacs,showkeys]{revtex4-1}
\usepackage{graphicx}
\usepackage{bm}
\usepackage{amsthm,amsfonts,amsmath,amssymb,amscd}

\begin{document}

%\preprint{v.1}

\title{Effect of interfacial Dzyaloshinskii-Moriya interaction on polarized neutrons reflection}


\author{D.~A.~Tatarskiy}
\affiliation{Institute for physics of microstructure of RAS, Nizhniy Novgorod, 603950, GSP-105, Russia}
\affiliation{Lobachevsky University, Gagarin ave., 23, Nizhniy Novgorod, 603950, Russia}
\email[]{tatarsky@ipmras.ru}


\date{\today}

\begin{abstract}
The antisymmetric Dzyaloshinskii-Moriya interaction (DMI) in noncentrosymmetric systems leads to various nonuniform chiral magnetic textures. Polarized neutron scattering is a powerful method for investigation such chiral distributions. The most used technique is a small-angle neutrons scattering (SANS). Multilayered magnetic films with the interface induced DMI (iDMI) can't be investigated by SANS because of their small volume. The appropriate technique is a polarized neutron reflectometry. Within the framework of the continuum theory of micromagnetics, we explore the impact of the iDMI on the polarized neutrons reflection from multilayers with random magnetic anisotropy. It is shown that the iDMI gives rise to a polarization-dependent asymmetric term in the reflection.
\end{abstract}


\maketitle

\section{\label{Intro}Introduction}
Recently magnetic multilayered structures emerge a great interest due to possible applications in magnetic memory and logic~\cite{Fert}. The fact is that the absence of the inversion symmetry in layered systems may lead to the antisymmetric exchange interaction~\cite{Dzyal,Moriya} on the ferromagnet-heavy metal interfaces: the interfacial induced Dzyaloshinskii-Moriya interaction (iDMI)~\cite{Crepieux,Anatomy,BergerInterlayer,Bogdanov}. The presence of the iDMI is probed by several methods: Brillouin light scattering~\cite{BLS1,BLS2,BLS3}, electron microscopy~\cite{LTEM,TEM1,TEM2,TEM3,TEM4,TEM5,TEM6,TEM7,TEM8,TEM9,TEM10,TEM11} etc. The polarized neutron reflectometry (PNR) is a powerful method for probing magnetic properties of different ferromagnetic films. Still it is not widely used for probing thin films with DMI~\cite{Monchesky}. The most appropriate method is small-angle neutron scattering for investigation materials with bulk DMI: noncentrosymmetric B20 crystals~\cite{B201,Muhl,Grig,B202,B203,B204,B205,B206,B207} or defect-induced DMI~\cite{MetlovTh,DMIMetlov}. An asymmetric term in SANS appears in case of the defect-induced DMI when the incident vector is perpendicular to the external field. The similar scattering is expected to be in the PNR experiment when the external field is applied along to the specular reflection plane perpendicularly to the multilayered film normal.

In this paper we predict an asymmetry in PNR from thin ferromagnet-heavy metal multilayered films with iDMI near saturation magnetization. The paper is organized as follows. In the first part we develop a continuum micromagnetic theory for small magnetization deviations near saturation in the presence of the iDMI. Than we calculate the differential scattering cross-section of polarized neutrons. At last part we estimate an experimental details for observations.

\begin{figure}[h!]
\includegraphics[width=0.6\linewidth,keepaspectratio=true]{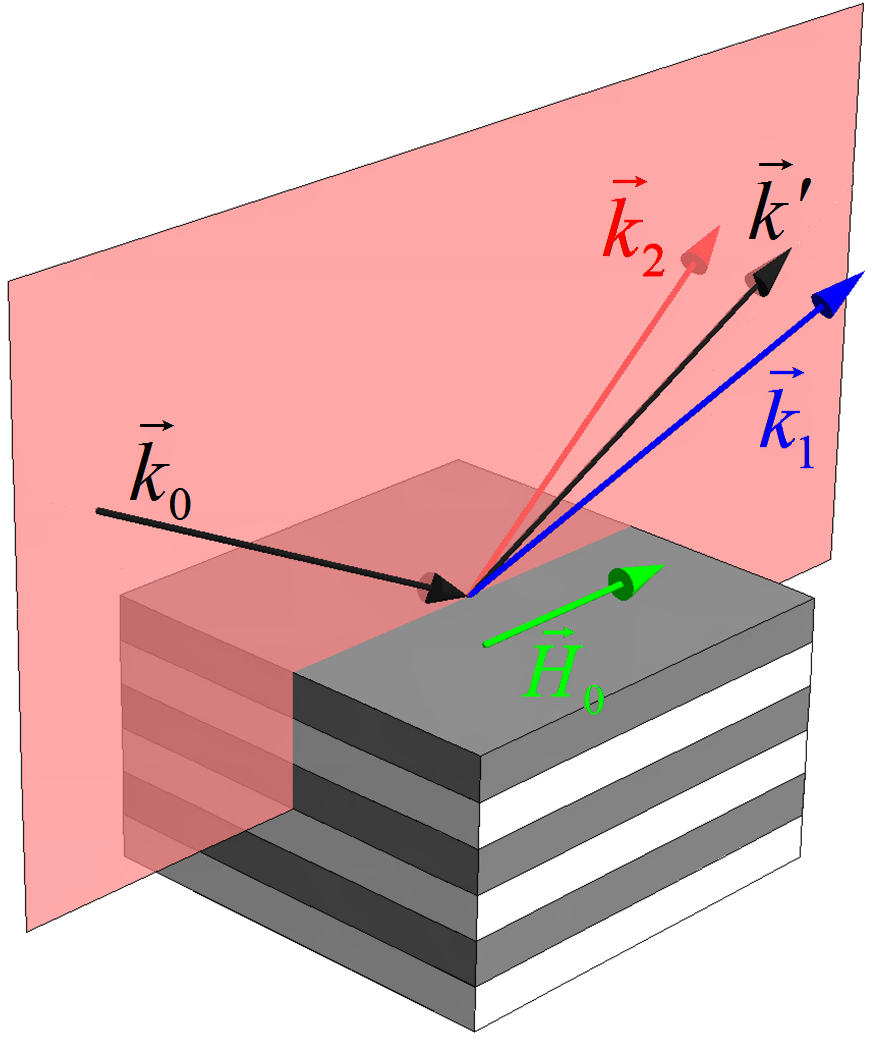}
\caption{\label{Fig1} (Color online) Geometry of reflection. The $\vec{k}_0$  is an incident wavevector, the $\vec{k}'$ specular reflection wavevector, the $\vec{k}_{1,2}$ denote directions of asymmerty for polarized neutron reflection.}
\end{figure}

\section{\label{Magn}Micromagnetic theory}
We present a micromagnetic model for small fluctuations of the magnetization in thin film with the iDMI near saturation. We have to solve the static equation to find this deviations~\cite{Brown,Aharoni,Kron}
\begin{equation}\label{Torques}
\left[\vec{M} \times \vec{H}^{eff} \right] = 0,
\end{equation}
where $\vec{H}^{eff}$ is an effective field. It is defined as the functional derivative of the magnetostatic energy-density functional $\vec{H}_{eff} = - \frac{\delta E}{\delta \vec{M}}$. This density consists of the exchange interaction, random anisotropy, Zeeman, magnetostatic and iDMI energies~\cite{MetlovTh}.
We consider the case when the external field is applied in plane of the film along $x$ axis. The $z$ axis is oriented along the normal to the film (Fig.~\ref{Fig1}). The magnetization is  written in the following form for these case
\begin{equation}\label{Magns}
\vec{M} = \left\{ M_0 , M_y \left( \vec{r} \right), M_z \left( \vec{r} \right) \right\}.
\end{equation}
We suppose that all the material characteristics are homogeneous in the real space. The anisotropy constant is homogeneous and only the anisotropy axis direction varies in the film.

Further we solve all the equations in the reciprocal space. The Fourier transform is given by the equation $\tilde{A} \left( \vec{q} \right) =\frac{1}{\left( 2 \pi \right)^3} \iiint {A \left( \vec{r} \right) \exp \left(- i \vec{q} \vec{r} \right) d^3 r}$. The algebraic operations in real space are substituted by convolution of the Fourier components $\tilde{A} \bigotimes \tilde{B} \left( \vec{k} \right) = \sum_{\vec{k_1}} \left( \tilde{A}\left( \vec{k_1} \right) \tilde{B} \left( \vec{k} - \vec{k}_1 \right) \right)$. The balance-of-torques equation~(\ref{Torques}) is solved in the first order of small deviations $M_{y,z}$. The effective field has the following exchange, magnetostatic, iDMI and random anisotropy components~\cite{Field,MetlovTh}
\begin{eqnarray}\label{Fields}
& \tilde{\vec{H}}_{ex}=- L_{ex}^2 q^2 \tilde{\vec{M}}, \nonumber \\
& \tilde{\vec{H}}_{ms}=-4 \pi \frac{\vec{q} \left(\vec{q} \cdot \tilde{\vec{M}}\right)}{q^2}, \\
& \tilde{\vec{H}}_{DMI}= -i L_{DMI} \Big( q_x \tilde{M}_z \vec{e}_x + q_y \tilde{M}_z \vec{e}_y - \nonumber \\
& \qquad \qquad \qquad \qquad- \left( q_x \tilde{M}_x + q_y \tilde{M}_y \right) \vec{e}_z \Big), \nonumber \\
& \tilde{\vec{H}}_{Ay,z}= K M_0 \sum_{\vec{k}}{n_x\left( \vec{k} \right) n_{y,z}\left( \vec{q}-\vec{k} \right)}, \nonumber
\end{eqnarray}
where $L_{ex}=\sqrt{\frac{2 J}{M_s^2}}$, $J$ is the exchange constant, $M_s$ is a saturation magnetization, $L_{DMI} = \frac{2 D}{M_s^2}$, $D$ is the iDMI constant,  $\vec{n}\left( \vec{q} \right)$ is a Fourier image of the local direction for random anisotropy axis. We obtain the system of linear equations with non-zero right part
\begin{equation}\label{LinearSys}
\begin{cases}
\left( A + 4 \pi \frac{q_y^2}{q^2} \right) \tilde{M}_y + B \tilde{M}_z =  H_{Ay},
\\
B^{*} \tilde{M}_y + \left( A + 4 \pi \frac{q_z^2}{q^2} \right) \tilde{M}_z = H_{Az},
\end{cases}
\end{equation}
where $A = \frac{H_0}{M_s} +  L_{ex}^2 q^2$ and $B = \left(4 \pi \frac{q_y q_z}{q^2}  + i q_y L_{DMI} \right)$. The solution of equations are
\begin{equation}\label{MSolution}
\begin{cases}
\tilde{M}_y = \frac{ H_{Ay} \left( A+4 \pi \frac{q^2_z}{q^2} \right) -  H_{Az} B}{A^2 + 4 \pi A \frac{q^2_y + q^2_z}{q^2} - q^2_y L_{DMI}^2},
\\
\tilde{M}_z = \frac{ H_{Az} \left( A+4 \pi \frac{q^2_y}{q^2} \right) -  H_{Ay} B^{*}}{A^2 + 4 \pi A \frac{q^2_y + q^2_z}{q^2} - q^2_y L_{DMI}^2}.
\end{cases}
\end{equation}

We assume the vector $\vec{H}_A$ takes all orientations with equal probabilities. Thus every value should be averaged over all angles of $\vec{H}_A = H_A\left( q \right) \left\{0,\cos \beta, \sin \beta\right\}$. Obviously the nonvanishing values will contain only $H_{Ay,z}^2$ amplitudes. They are proportional to the isotropic $H_A^2\left( q \right)$ correlation. The anisotropy connected with structural characteristics of the polycrystalline ferromagnetic film and can be written in the following form
\begin{equation}\label{Anisotropy}
H_A^2\left( q \right)= \frac{\langle H_A^2 \rangle}{\left(1+\xi^2 q^2\right)^2},
\end{equation}
where $\langle H_A^2 \rangle$ is the mean-square anisotropy field and $\xi$ denotes the correlation length . For an idealized nanocrystalline ferromagnet, where each grain is a single crystal and the anisotropy field jumps randomly in direction at grain boundaries due to the changing set of crystallographic axes, the correlation length $\xi$ is expected to be related to the average grain size.

\section{\label{Scat}Polarized neutrons grazing incidence reflection}

The interaction of the thermal neutrons with condensed matter is described by the following Schr\"{o}dinger equation
\begin{equation}\label{Shrod}
\hat{H} = \frac{\vec{p}^2}{2 m} + V \left( \vec{r} \right) + \mu \left( \hat{\vec{\sigma}} \cdot \vec{B} \left( \vec{r} \right) \right),
\end{equation}
where $m$ and $\mu$ are mass and magnetic moment of neutron, $V \left( \vec{r} \right)$ is scalar nuclear potential and $\vec{B} \left( \vec{r} \right)$ is the magnetic induction. The differential elastic  scattering cross section of polarized neutrons has the following form in the first order of the perturbation theory
\begin{widetext}
\begin{equation}\label{DCSPN}
\frac{\partial\sigma\left(\vec{q},\vec{P}\right)}{\partial\Omega}= \left(\frac{m}{2 \pi \hbar^2}\right)^2 \bigg( \left|\tilde{V}\right|^2+\mu^2 \left|\tilde{\vec{B}}\right|^2+\mu \left(\vec{P}\cdot \left(\tilde{V}^{*}\tilde{\vec{B}}+\tilde{V}\tilde{\vec{B}}^{*}\right)\right) + i\mu^2 \left( \vec{P} \cdot \left[ \tilde{\vec{B}}^{*} \times \tilde{\vec{B}} \right] \right) \bigg),
\end{equation}
\end{widetext}
where $\vec{q} = k_0 \left(
  \begin{array}{c}
    - \frac{\gamma^2}{2} \cos \alpha \\
    \gamma \cos \alpha \\
    2 \sin \alpha \\
  \end{array}
\right)$ is the momentum transfer wavevector, $k_0$ is the incident neutrons wavevector, $\alpha$ is the glancing angle, $\gamma$ is the transversal deviation angle from the specular reflection plane and $\left|\vec{P}\right| \leq 1$ is the average polarization vector of the neutron beam. The magnetic induction vector $\vec{B}$ can be derived from magnetostatic problem solution. In the Fourier space it has the following form

\begin{equation}\label{MSSolution}
\tilde{\vec{B}} = \left(\vec{H}_0 + 4 \pi M_0 \vec{e}_x\right) \delta \left( \vec{q} \right) -4 \pi \frac{\left[ \vec{q} \times \left[ \vec{q} \times \tilde{\vec{M}} \right] \right]}{q^2}.
\end{equation}

Substituting~(\ref{MSSolution}) in~(\ref{DCSPN}) and doing some vector transformations we obtain the form of the last term in~(\ref{DCSPN})
\begin{equation}\label{Symm}
\frac{\partial\sigma\left(\vec{q},\vec{P}\right)}{\partial\Omega} = i \left(\frac{2 m \mu}{\hbar^2}\right)^2 \frac{\left(\vec{P} \cdot \vec{q}\right)}{q^2} \left( \vec{q} \cdot \left[ \tilde{\vec{M}}^* \times \tilde{\vec{M}} \right] \right).
\end{equation}
This term is non-zero only in case of non-zero image part of the magnetization deviations~(\ref{MSolution}). Thus the iDMI is the necessary condition for asymmetry reflection observation.

Because of a weak neutron scattering from the single interface we consider multilayered periodical film to get strong enough reflectivity near Bragg conditions. The geometry of reflection is given on the Fig.~\ref{Fig1}. The asymmetry of scattering is perpendicular to the reflection plane. So the reflections differ in case of $\vec{k}_1$ or $\vec{k}_2$ scattering. We assume the nonmagnetic interfaces are thinner than ferromagnetic layer and we can omit spatial variations of exchange, anisotropy and iDMI through the periodic structure depth. Thus we can substitute solution~(\ref{MSolution}) into the differntial scattering cross-section term~(\ref{Symm}). Also we omit the terms of the higher orders in $\gamma$. The final result for asymmetrical part of scattering is following
\begin{widetext}
\begin{equation}{\label{Answer}}
\frac{\partial\sigma\left(q,\alpha,\gamma,P \right)}{\partial\Omega} = k_0 L_{DMI} P  \left(\frac{2 m \mu}{\hbar^2}\right)^2 \gamma^5 \cot \alpha  \frac{\langle H_A^2 \rangle}{\left(1+ \xi^2 q^2\right)^2} \frac{\left( 2 A+4 \pi \right)}{A^2 \left( A + 4 \pi \right)^2}.
\end{equation}
\end{widetext}

\section{\label{Coda}Discussion}

\begin{table}
\caption{\label{Constants}Magnetic and structural parameters of Co/Pt multilayers and neutron beam charachteristics are used for estimations~\cite{BLS3,Komogortsev}.}
\begin{ruledtabular}
\begin{tabular}{c c}
 $M_s$ (erg Gs cm$^{-1}$) & 1000 \\
 $H_0$ (Oe) & 1000--3000 \\
 J ($\mu$erg/cm) & 0.5 \\
 D (erg/cm$^2$) & 1--3 \\
 $\lambda$ (nm) & 0.3 \\
 $k_0$ (nm$^{-1}$) & 20 \\
 $q$ (nm$^{-1}$) & 2 \\
 $L_{ex}$ (nm) & 10 \\
 $L_{DMI}$ (nm) & 10 \\
 $\alpha$ (mrad) & 50 \\
 $\xi$ (nm) & 5--10 \\
\end{tabular}
\end{ruledtabular}
\end{table}

We obtain estimations for~(\ref{Answer}) substituting material constants for Co/Pt multilayered film with period $d \approx 3$ nm and neutron beam with wavelength 0.3 nm near Bragg conditions for such structure. For fully polarized neutron beam $P=1$ and substitution parameters from table~\ref{Constants} to~(\ref{Answer}) gives the result
\begin{equation}{\label{Answer2}}
\frac{\partial\sigma\left(\gamma\right)}{\partial\Omega} \approx 10^{6} \langle H_A^2 \rangle \gamma^5 = 10^{-4}\langle H_A^2 \rangle.
\end{equation}
This equation holds for small $\gamma \approx 1-10 $ mrad.

Comparing~\cite{DMIMetlov} we see the great difference in chiral fluctuation for small transversal magnetization deviations in case of the bulk DMI and the interfacial DMI. In bulk materials the chiral part of deviations is a long range helix-like fluctuations, e.g. $m_{x,y} \sim \pm i L_{DMI} q_z$. Thus the asymmetry in polarized neutron scattering exists when the external field applied along this helix axis and perpendicular to the scattering wavevector. The iDMI give rise to the long range cycloidal-like fluctuations, $m_{y,z} \sim \pm i L_{DMI} q_y$. That is why the asymmetry effect appears when the external field applied in the specular reflection plane.

As it follows from~(\ref{Answer2}) the crystal magnetic anisotropy correlations $\langle H_A^2 \rangle$ and iDMI strength $L_{DMI}$ can't be measured by such neutron asymmetry scattering separately. Still if one can somehow ``fix'' the polycrystal grain sizes and tune iDMI, e.g. straining the media~\cite{Strain,DMIMetlov} or applying currents~\cite{Bias1,Bias2}, the neutron reflection asymmetry will give us the oportunity to measure $L_{DMI}$.

\begin{acknowledgments}
Authors are grateful to RFBR Grant \#18-32-20036 for financial support and to O.~G.~Udalov and A.~A.~Fraerman for fruitful discussions.
\end{acknowledgments}

\bibliography{3rdOrderNeutrons}

\end{document}